\documentclass[11pt]{article}
\usepackage[preprint]{nlp4call} 
\usepackage{times}
\usepackage{xurl}
\usepackage{latexsym}

\usepackage{amsmath,amssymb,amsfonts, comment}
\usepackage{algorithmic}
\usepackage{graphicx}
\usepackage{textcomp}
\usepackage{xcolor, url, subfigure, subcaption, multicol, multirow, booktabs, bm, enumitem, tabularx, makecell}
\usepackage{tikz}
\usetikzlibrary{calc}
\usepackage[table]{xcolor}

\newcommand\blfootnote[1]{%
  \begingroup
  \renewcommand\thefootnote{}\footnote{#1}%
  \addtocounter{footnote}{-1}%
  \endgroup
}

\title{Data Augmentation for L2 English Speaking Assessment using TTS}

 \author{Stefano Bannò, Penny Karanasou, Mengjie Qian, Kate M. Knill, Mark J. F. Gales \\
 ALTA Institute\\
  Machine Intelligence Laboratory, Department of Engineering \\
  University of Cambridge, UK \\
  \texttt{\{sb2549,pk407,mq227,kmk1001,mjfg1000\}@cam.ac.uk}}

\date{}

\begin{document}
\maketitle

\blfootnote{Demo examples: \url{sbanno.github.io/l2-synthetic-speech/}}

\begin{abstract}
Automated assessment of second language (L2) speaking proficiency requires substantial annotated speech data, which are scarce compared to written learner corpora. We investigate whether written L2 responses can be transformed into useful synthetic speech for proficiency assessment using text-to-speech (TTS) and voice cloning. Using COREFL, a corpus of paired spoken and written responses from L2 learners of English, we systematically study two factors: how written responses should be transformed into spoken-style language (``speechification'') and how synthetic voices and texts should be paired based on shared learner attributes (proficiency level, first language, both, or neither). We generate speechified responses with a large language model and synthesise them using TTS and voice cloning, then evaluate their utility for audio-based (HuBERT) and text-based (ModernBERT) proficiency grading. Results show that pairing voices and texts based on proficiency provides a principled strategy for synthetic data generation, while speechification substantially improves the match between written and spoken L2 and improves downstream grading performance. Augmenting real training data with synthetic speechified responses improves both HuBERT- and ModernBERT-based graders, demonstrating the potential of synthetic spoken data for L2 speaking proficiency assessment.

\end{abstract}

\section{Introduction}
\label{sec:intro}

Automated assessment of second language (L2) speaking proficiency is increasingly used in computer-assisted language learning systems, providing scalable and consistent scoring of learner performance~\cite{zechner2019automated}. Despite these advances, progress in automated L2 speaking assessment research remains constrained by the limited availability of large-scale, well-annotated spontaneous speech corpora. While large L2 written corpora are widely available~\cite{geertzen2013automatic, blanchard2013toefl11, nicholls2024write, crossley2023english}, spoken datasets remain comparatively scarce due to the cost of recording, transcription, and expert annotation. 

A promising way to address this imbalance is the use of zero-shot text-to-speech (TTS) and voice-cloning technologies. These frameworks enable the conversion of abundant, diverse, and easily accessible L2 written essays into synthetic speech. Recent studies have started to investigate the use of TTS synthesis to augment data for L2 spontaneous speech assessment. In \citet{wang2025novel}, proficiency-conditioned responses are generated with a large language model (LLM) and converted into speech using speaker-conditioned TTS to augment training data for automated speaking assessment. Whilst improvements are reported in downstream scoring performance through fine-tuning a multimodal LLM-based grader, the quality of the generated speech itself is not analysed (e.g., speaker similarity, transcription quality, or linguistic realism), and the gains over a BERT-based~\cite{devlin2018} baseline remain relatively modest. Similarly, in \citet{voskoboinik2025enhancing}, the use of LLM-generated synthetic transcripts is explored for automated speaking assessment in low-resource languages. Using GPT-4~\cite{openai2023gpt4} to generate proficiency-conditioned learner responses, it shows that transcript-level data augmentation can improve scoring performance. However, this approach operates exclusively on transcripts without including TTS. In \citet{karanasou25_slate}, a more detailed analysis of acoustic and content-related aspects of the synthetic learner speech is provided, but the primary focus is spoken grammatical error correction rather than assessment. Additionally, the approach does not rely on essays, but on manual transcriptions, which remain costly to obtain. 


In this study, by contrast, we aim to leverage essays to augment L2 learner spoken data for proficiency assessment. A fundamental challenge remains: the disparity between written and spoken language. Spontaneous L2 speech is defined by online speech-planning phenomena, resulting in disfluencies such as hesitations, self-repairs, and repetitions, and discourse markers that are tightly linked with learners' oral proficiency~\cite{huang2023development, yan2025disfluency}. Conversely, written essays reflect a deliberate planning process characterised by higher syntactic complexity and greater lexical richness~\cite{biber2002speaking}. 

This raises the question of what is actually required to generate synthetic L2 speech that is suitable for assessment purposes, and more broadly, how written L2 data can be reliably transformed into spoken-like data. Addressing this question requires analysing the relationship between written and spoken L2 productions. The few existing L2 speech corpora provide valuable resources in isolation, but they are not designed for this purpose. ICNALE~\cite{ishikawa2011}, CLES~\cite{coulange-etal-2024-corpus}, Speechocean762~\cite{speechocean762}, and L2-ARCTIC~\cite{zhao2018l2} are limited either in first language (L1) diversity, modality (read rather than spontaneous speech), or proficiency level coverage. The Speak \& Improve Corpus~\cite{sicorpus25, qian25_challenge} represents a major milestone, providing large-scale spontaneous L2 speech with detailed proficiency annotations, but its speakers, tasks, and exam parts are not directly comparable to its written counterpart, the Write \& Improve Corpus~\cite{nicholls2024write}. 

In this work, we use the Corpus of English as a Foreign Language (COREFL)~\cite{lozano2020designing} (see Section \ref{sec:data}). What differentiates this corpus from the above-mentioned ones is that it contains paired written and spoken answers to the same questions from the same L2 learners. 
This double modality allows us to study how linguistic and speaker-level factors interact in synthetic speech generation, enabling controlled investigation of different speaker–text pairing strategies as discussed below.


In our proposed data augmentation framework, to address the structural differences between modalities, we first introduce a \emph{text speechification} step using a large language model (LLM). This module rewrites written responses into spoken-style transcripts. Next, the resulting texts are used for audio generation with a TTS/voice-cloning model. To select the voice for the generated audio, we investigate different matching strategies for pairing learner voices and written texts based on shared learner attributes (proficiency level, L1, both, or neither). Ensuring consistency between the textual and acoustic dimensions is important because, in real-world scenarios, the downstream grading system is not known a priori and may be text-based, audio-based, or multimodal. Finally, we validate our augmented data by evaluating downstream performance on two automated grading models: an audio-based system using a self-supervised learning (SSL) speech foundation model, i.e., HuBERT~\cite{hsu2021hubert}, and a text-based system using a text foundation model, i.e., ModernBERT~\cite{warner-etal-2025-smarter}. Both SSL-based~\cite{xu21, peng21, eesung, banno23_slate, mcknight23_slate} and BERT-like~\cite{craighead2020, raina2020universal, wang2021} architectures have been extensively used in L2 speaking assessment. The main contributions of this paper are as follows:

\begin{itemize}[leftmargin=1em]
    \item we introduce an LLM-based preprocessing step that converts written responses into spoken-style transcripts, reducing the discrepancy between written and spoken language before TTS generation;
 
    \item we study how to construct high-quality synthetic L2 speech for speaking assessment by pairing speakers and texts according to proficiency level and L1, finding that proficiency-level matching provides a principled strategy for synthetic data augmentation that is consistent across downstream assessment models;


    \item we show that our data augmentation approach consistently improves performance across both text-based and audio-based grading systems.
\end{itemize}

\section{Data}
\label{sec:data}

For our experiments, we use the Corpus of English as a Foreign Language (COREFL)~\cite{lozano2020designing}, a publicly available dataset\footnote{\url{corefl.learnercorpora.com/}} that contains both written and spoken answers to the same
questions by L2 learners of English from 9 different L1 backgrounds, with German and Spanish representing the highest percentage of learners, followed by Greek, Italian, Chinese, Turkish, Czech, Estonian, and French learners. However, only the first 5 L1s are also represented in the spoken section.

A unique feature of this corpus is that some learners were asked to produce both a spoken and written response to the same question with a few weeks of difference between the two tasks. The corpus includes four question prompts. Learners are asked to: describe the picture story \emph{Frog, where are you?} \cite{mayer2003frog}; talk about a famous person; describe a film they have seen recently; describe an excerpt from Charlie Chaplin's \emph{The Kid}.

Although the spoken section of the corpus is relatively small, it contains comparatively long responses: they range from approximately 40 seconds to 8 minutes with an average duration of 3 minutes. All the spoken responses are accompanied by manual transcriptions. The written responses range from roughly 10 to 1,000 words, with an average of 160 words.
Table \ref{tab:data} shows the speaker-independent training and test set splits.
All learners included in the test set also produced a written response, whereas not all learners included in the training set also have a written counterpart.
Additionally, there are 2,820 written responses that are not accompanied by their spoken counterparts. We refer to this set as \emph{Extra}.

\begin{figure}[ht!]
    \centering
        \includegraphics[width=0.8\linewidth]{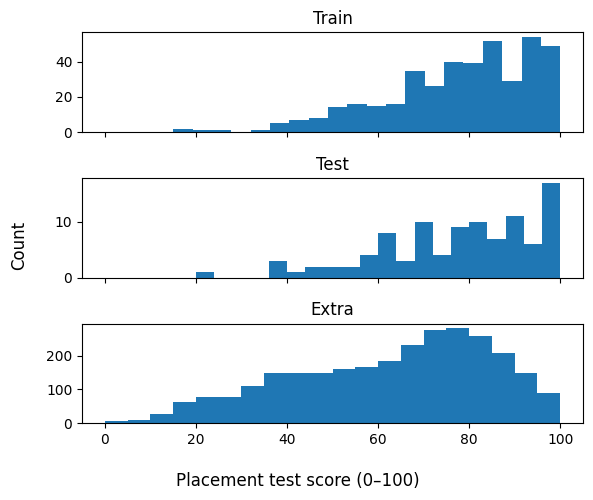}

    \caption{Placement test scores (0-100) distribution across sets.}
    \label{fig:histo}
\end{figure}

The corpus is annotated with proficiency levels based on the Common European Framework of Reference (CEFR)~\cite{cefr2001}. These were given to the learners after they took a placement test, which assigned them a score ranging from 0 to 100 (i.e., A1 to C2). We normalise this score to [0, 1] and use it as the target variable for model training in our language assessment experiments.
There is also a control L1 English subcorpus. For our experiments, we use this subcorpus as additional C2 data. Figure \ref{fig:histo} shows the placement test score distribution.

\begin{table}[ht]
\small
\centering

\begin{tabular}{lrr|rr}
\hline
& \multicolumn{2}{c|}{\textbf{Spoken}} &
\multicolumn{2}{c}{\textbf{Written}} \\
\cline{2-5}
\textbf{Set} &
\textbf{Responses} &
\textbf{Hours} &
\textbf{Responses} &
\textbf{Words} \\
\hline
Train & 410 & 16.7 & 383 & 111K \\
Test  & 100 & 4.2 & 100 & 27K \\
\hline
Extra & -- & -- & 2,820 & 765K \\
\hline
\end{tabular}
\caption{Composition of COREFL (spoken and written).}
\label{tab:data}
\end{table}

\section{Data Augmentation Framework}


We propose a data augmentation framework that leverages L2 learner texts to generate synthetic spoken responses for the automated assessment of holistic speaking proficiency. As outlined in Section \ref{sec:intro}, it consists of two phases: LLM-based speechification and attribute-conditioned speaker-text matching and synthesis.

\subsection{Text ``speechification''}

\label{sec:speechify}

To improve the quality of synthetic responses, we introduce a \emph{text speechification} step. Given a learner's written response and CEFR level, we use an LLM to transform the essay-like text into a spoken-style learner transcript. The prompt, reported in Appendix \ref{sec:appendix_speechify}, instructs the model to preserve the original meaning and all learner errors while rewriting the response as spontaneous speech. Specifically, the generated transcript incorporates characteristics of natural spoken production, including incremental discourse structure, hesitations, repetitions, self-repairs, false starts, and occasional vague references. The frequency and complexity of these phenomena are conditioned on the learner's CEFR level, with lower-proficiency learners exhibiting more disfluencies and simpler discourse structures. The resulting transcript is intended to approximate the characteristics of authentic learner speech while remaining faithful to the content of the original written response.

We also include a control experiment in which we remove CEFR-conditioning from the prompt. This can be found in Appendix \ref{sec:appendix_speechify}.

\subsection{Speaker-text matching strategies for synthesis}
\label{sec:data_aug}

We refer to real spoken data as \emph{Real$_{\tt Spoken}$} and to real written data as \emph{Real$_{\tt Written}$}. In the following, we describe the matching conditions used to generate synthetic data. First, we synthesise a learner's spoken response by using its own manual transcription and the same learner's voice. We refer to this condition as \emph{Synth$_{\tt Trs}$}. This serves as a best-case reference for evaluating the TTS system, as both the speaker identity and linguistic content from the same spoken response are preserved. We then synthesise a learner's written production using the same learner's voice, referred to as \emph{Matched$_{\tt Spkr}$}. In this case, although the information is produced by the same learner, the linguistic content differs between the written and spoken productions.
Subsequently, we synthesise spoken data from written data using different pairing strategies to investigate how conditioning on learner attributes affects the similarity between synthetic and real speech. For each learner voice, a written response is chosen and assigned to it. A voice may match the selected written response with respect to CEFR level (\emph{Matched$_{\tt CEFR}$}), L1 (\emph{Matched$_{\tt L1}$}), both attributes (\emph{Matched$_{\tt CEFR\&L1}$}), or neither (\emph{Random}). When no eligible text satisfies the primary matching criterion, we progressively relax the constraints. For \emph{Matched$_{\tt CEFR\&L1}$}, the fallback order is exact CEFR+L1 and adjacent CEFR level with the same L1. For \emph{Matched$_{\tt CEFR}$}, the order is exact CEFR and adjacent CEFR level. For \emph{Matched$_{\tt L1}$}, the order is exact L1 followed by an unrestricted global pool. In all cases, except \emph{Synth$_{\tt Trs}$} and \emph{Matched$_{\tt Spkr}$}, the voice's own written responses are excluded from the candidate pool. To promote diversity in the synthetic datasets, we use a balanced selection mechanism which tracks the frequency with which voices and texts are selected and avoids repetition of speakers or responses.
To improve robustness, we generate five different draws for each matching condition. Each draw corresponds to a different random assignment of voices to texts under the same matching constraints, such that a voice paired with one text in a given draw may be paired with a different text in another.\footnote{With respect to the fallback system, the \emph{Matched$_{\tt CEFR\&L1}$} test set only has 5 adjacent cases for each draw, while the \emph{Matched$_{\tt CEFR}$} test set only has 1 adjacent case for each draw. The \emph{Matched$_{\tt L1}$} and \emph{Random} test sets fully satisfy their criteria.}

\section{Experimental Setup}
\label{sec:experimental_setup}

The first objective of this work is to investigate the requirements and criteria for generating synthetic data suitable for L2 spontaneous speaking assessment. To this end, we first synthesise the test set data (see Section \ref{sec:exp_test}). The underlying idea is that, if an automated grading system trained on real data produces similar results when tested on synthetic and real audio data, then the synthetic data are likely to preserve the characteristics that are relevant for assessment.

Based on this initial analysis, we then revisit the synthetic generation pipeline. In particular, after observing that synthetic speech generated directly from raw written responses leads to a substantial degradation in the text-based grading system, we introduce a text speechification step (see Section \ref{sec:exp_speechify}) to transform written responses into spoken-style transcripts before TTS synthesis. We subsequently evaluate whether this modification improves the quality of the generated speech.

Finally, having identified both the most effective speaker-text pairing strategy and the benefit of speechification, we apply the resulting pipeline to the training data (see Section \ref{sec:exp_training}). The generated synthetic responses are used for data augmentation to investigate whether additional synthetic training examples improve the performance of our graders.

\subsection{Models}
\label{sec:models}

\subsubsection{TTS} As our TTS/voice-cloning system for spoken data augmentation, we use OmniVoice~\cite{zhu2026omnivoiceomnilingualzeroshottexttospeech},\footnote{\url{github.com/k2-fsa/OmniVoice}}
a zero-shot multilingual TTS model supporting over 600 languages, based on a non-autoregressive diffusion language model architecture that directly maps text to multi-codebook acoustic tokens.
To generate synthetic spoken responses, the written responses in the test set are first segmented into sentences using the \texttt{spaCy}\footnote{\url{spacy.io}} \emph{sentencizer}. Voice cloning is then applied at the sentence-level using the first 10 seconds of the reference audio, and the resulting synthetic utterances are concatenated to reconstruct the full spoken response.

\subsubsection{Automatic speech recognition (ASR)} In a real-world deployment, manually transcribing every learner's spoken response is infeasible. Consequently, text-based grading systems must rely on ASR to produce the transcriptions used for assessment. To do this, we use the publicly available CrisperWhisper~\cite{zusag24_interspeech},\footnote{\url{github.com/nyrahealth/CrisperWhisper}}
a variant of OpenAI’s Whisper~\cite{radford2023robust} designed for verbatim speech recognition. Unlike the original Whisper model, which tends to normalise speech and omit disfluencies in favour of a more fluent transcription style, CrisperWhisper aims to transcribe speech as accurately as possible, preserving all spoken content, including fillers and disfluencies.

\subsubsection{Grading systems}
\label{sec:grading_sys}
For automatic language assessment, i.e., the main task of this paper, we employ an audio-based and a text-based grading system. The former is based on HuBERT~\cite{hsu2021hubert},\footnote{Initialised from \texttt{facebook/hubert-base-ls960}.} with an architecture similar to that described in \citet{mcknight23_slate}, which uses wav2vec2~\cite{baevski}.
The audio recordings are first segmented into overlapping fixed-length chunks prior to modelling. We use 15-second segments with a 3-second overlap to ensure coverage of long-form responses while preserving local temporal context. At model level, HuBERT is used as a frozen feature encoder, and we introduce a regression head operating on pooled hidden representations. Instead of simple pooling, we apply a hierarchical attention-based aggregation mechanism over frame-level encoder outputs. This mechanism learns to weight temporal frames and segment-level representations before producing a final utterance-level embedding, which is passed through a feed-forward regression head to predict the score.



The text-based grading system is based on ModernBERT~\cite{warner-etal-2025-smarter}.\footnote{Initialised from \texttt{answerdotai/ModernBERT-base}.} The model is fine-tuned using a standard sequence regression head.

For both grading systems, we use an ensemble of two graders using two different random seeds, and training minimises mean squared error between predicted and placement scores normalised from 0 to 1 (see Section \ref{sec:data}) for a maximum of 20 epochs. The training hyperparameters are reported in Table \ref{tab:hyperparameters}.

\begin{table}[!ht]
\small
\centering
\begin{tabular}{lcc}
\hline
\textbf{Hyperparameter} & \textbf{HuBERT} & \textbf{MBERT} \\
\hline
Learning rate &  $1 \times 10^{-5}$  & $2 \times 10^{-5}$  \\
Batch size & 8  & 32  \\
Gradient accumulation steps & 4  & -  \\
Early stopping patience & 3  & 3  \\
\hline
\end{tabular}
\caption{Grading systems training hyperparameters.}
\label{tab:hyperparameters}
\end{table}

For each of the two training seeds, we randomly select 10\% of the real spoken training data (41 samples) as a development set for optimisation. To avoid data contamination, the candidate IDs selected for the development set in each seed are excluded from that seed's augmented training sets, including both their written texts and speech recordings.

As will be illustrated in Section \ref{sec:exp_training}, our best-performing system uses a training set composed of real learner data and synthetic data generated by matching learners' texts and voices based on CEFR level. This means that, for the synthetic section of the training data, we have two possible target training labels: one placement score referred to voices and one associated with texts. However, since they are matched based on CEFR level, their differences are tiny, i.e., the mean of the absolute difference between the two sets of labels on the whole dataset is 0.043 and their standard deviation is 0.034. Therefore, we chose the labels associated with voices as training labels.

\subsubsection{Text speechifier} To transform the original written responses into spoken-style learner transcripts, we use GPT-4.1~\cite{openai2023gpt4}. The cost of speechifying the 383 written responses in the training set was approximately \$2.5. Generating speechified versions of the 2,820 responses in the \emph{Extra} set (see Section \ref{sec:exp_training}) incurred an additional cost of approximately \$15.

\subsection{Evaluation metrics}
\label{sec:evaluation_metrics}

\subsubsection{Speaker verification} To assess whether speaker identity is preserved after voice cloning, we compute cosine similarity between the speaker embeddings of the real and synthetic speech. The speaker embeddings are extracted using \texttt{pyannote}~\cite{Bredin2020}.

\subsubsection{ASR evaluation} We use CrisperWhisper to transcribe both real and synthetic spoken responses. As these transcripts are used in the text-based grading system, transcription quality is critical. We evaluate them using word error rate (WER).

\subsubsection{Speech quality}

To assess the perceptual quality of the synthetic speech, we use UTMOSv2~\cite{baba2024utmosv2}, a non-intrusive speech quality assessment model that predicts human ratings of speech quality from the audio signal. We report the mean UTMOS score across the generated responses ranging from 1 to 5, with higher scores indicating better perceived speech quality. This metric provides a complementary assessment to speaker verification, allowing us to evaluate not only whether the cloned voice preserves speaker identity, but also the overall quality and naturalness of the synthesised speech.

\subsubsection{Grading performance} To evaluate grading performance, we report Pearson’s correlation coefficient (PCC), Spearman’s rank correlation coefficient (SRC), and root mean square error (RMSE) between the reference and predicted scores.
For PCC and SRC, the higher value, the better. For RMSE, the lower value the better.

\subsubsection{Text analysis} \label{sec:txt_anl} We analyse textual differences between real and synthetic transcriptions using five metrics: average text length (Txt. len.), the average mean dependency distance (MDD)~\cite{liu2008dependency} as computed with \texttt{textdescriptives}\footnote{\url{github.com/HLasse/textdescriptives}}, hesitation ratio (Hes. rt.), Moving Average Type-Token Ratio (MATTR)~\cite{Covington2010} with a window size of 50 tokens, and the average number of unique difficult words (Dif. wds.) as computed with \texttt{textstat}.\footnote{\url{textstat.org/}}

\subsubsection{Text similarity} As described in Section~\ref{sec:speechify}, we speechify written responses. To ensure that this transformation preserves the original semantic content, we measure text similarity between the written responses before and after speechification using cosine similarity between sentence embeddings of original and speechified written responses. These are extracted using \texttt{all-mpnet-base-v2}~\cite{song2020mpnet, reimers-gurevych-2019-sentence}.

\section{Experimental results}

\subsection{Analysis and results on augmented test set without training}
\label{sec:exp_test}

As illustrated in Section~\ref{sec:experimental_setup}, we first evaluate the quality of the synthetic spoken responses on the test set across all matching conditions, prior to introducing speechification.

\begin{figure}[ht!]
    \centering
        \includegraphics[width=0.75\linewidth]{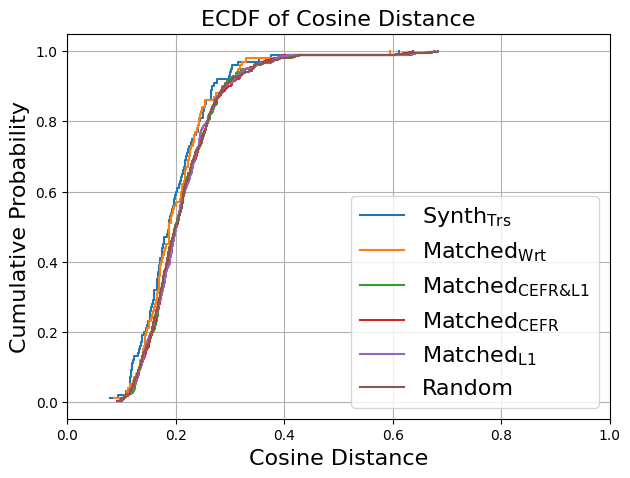}

    \caption{ECDF of cosine distances across matching conditions.}
    \label{fig:ecdf_sp}
\end{figure}

First, we evaluate whether speaker identity is preserved.
As illustrated in Section \ref{sec:evaluation_metrics}, we report the cosine similarity between the speaker embeddings of the real and synthetic speech for each matching condition in Table \ref{tab:results_WER}. Additionally, following \citet{karanasou25_slate}, we report the empirical cumulative distribution function (ECDF) of the cosine distances between speaker embeddings (i.e., real vs. synthetic) in Figure~\ref{fig:ecdf_sp}. 
As can be observed, no substantial differences are observed across matching conditions, except for \emph{Synth$_{\tt Trs}$} and \emph{Matched$_{\tt Spkr}$}, which yield the highest cosine similarity. This is expected, as we match learners' voices with their own manual transcriptions for \emph{Synth$_{\tt Trs}$} and with their own written production for \emph{Matched$_{\tt Spkr}$}. 
The few failures were traced to a recording with prolonged background noise before speech onset, which corrupted the speaker representation used for voice cloning.

We then compute WER on the CrisperWhisper transcriptions across the different matching conditions. The results are also reported in Table~\ref{tab:results_WER}.
We do not observe systematic differences in WER across the various conditions.

\begin{table}[ht!]
\small
\centering
\setlength{\tabcolsep}{2.8pt} 
\begin{tabular}{l|l|c|c}
\hline
\makecell{\textbf{Linguistic}\\\textbf{content source}} 
& \textbf{Test Set} & \textbf{WER} & \textbf{Cos. Sim.} \\
\hline

\multirow{2}{*}{Spoken}
& Real$_{\tt Spoken}$ & 11.98 & -- \\
& Synth$_{\tt Trs}$ & 11.03 & 0.803 \\
\hline\hline

\multirow{5}{*}{Written}
& Matched$_{\tt Spkr}$ & 12.00 & 0.797 \\
\cline{2-4}
& Matched$_{\tt CEFR \& L1}$ & $9.60_{\pm{1.45}}$ & $0.790_{\pm{.001}}$ \\
& Matched$_{\tt CEFR}$ & $9.87_{\pm{1.16}}$ & $0.789_{\pm{.001}}$ \\
& Matched$_{\tt L1}$ & $10.77_{\pm{2.10}}$ & $0.788_{\pm{.002}}$ \\
& Random & $10.87_{\pm{1.46}}$ & $0.788_{\pm{.002}}$ \\
\hline

\end{tabular}
\caption{WER and cosine similarity (Cos. sim.) across matching conditions. Mean and standard deviation are computed across the five draws.}
\label{tab:results_WER}
\end{table}

We can now use UTMOSv2 to score the speech quality of the TTS-generated data. Table \ref{tab:utmosv2} reports the results across matching conditions.
\begin{table}[ht]
\small
    \centering
    \begin{tabular}{l|l|c}
        \hline
        \makecell{\textbf{Linguistic source} \\ \textbf{content}} & \textbf{Test set} & \textbf{UTMOS} \\
        \hline
        \multirow{2}{*}{Spoken}
            & Real$_{\tt Spoken}$ & 2.205 \\
            & Synth$_{\tt Trs}$ & 2.326 \\
        \hline
        \multirow{5}{*}{Written}
            & Matched$_{\tt Spkr}$ & 2.366 \\
            \cline{2-3}
            & Matched$_{\tt CEFR\&L1}$ & $2.380_{\pm{.04}}$ \\
            & Matched$_{\tt CEFR}$ & $2.398_{\pm{.05}}$ \\
            & Matched$_{\tt L1}$ & $2.394_{\pm{.03}}$ \\
            & Random & $2.402_{\pm{.05}}$ \\
        \hline
    \end{tabular}
    \caption{Mean UTMOS scores results across matching conditions. Mean and standard deviation are computed across the five draws.}
    \label{tab:utmosv2}
\end{table}
UTMOSv2 is trained primarily on corpora of L1-speaker and professionally-recorded TTS output, and it is calibrated to treat deviations from fluent, native-like prosody and pronunciation as indicators of lower quality. As our reference recordings consist of authentic L2 English speech, they naturally contain non-native prosodic patterns, disfluencies, and accented pronunciation that are typical and expected features of L2 speech rather than genuine defects. As a result, UTMOS may penalise reference recordings for characteristics that are inherent to L2 speech rather than indicative of poor recording or speech quality. This can lead to artificially lower scores for the reference recordings compared with the synthetic systems. Such domain mismatch suggests that UTMOS scores should be interpreted cautiously when applied to L2 speech data. However, if we treat it as a comparative metric, all systems are shown to obtain similar results, with the lowest performance observed in \emph{Real$_{\tt Spoken}$}, suggesting that TTS synthesis may smooth over some of the L2-specific characteristics that would otherwise lower UTMOS scores.

We can then evaluate the performance of the grading systems, as shown in Tables \ref{tab:results_baselines} and \ref{tab:results_all}. It is important to note that, at this stage, neither grader has been exposed to synthetic data during training.
The HuBERT-based grader is trained solely on the real audio recordings, whereas the ModernBERT-based grader is trained solely on the corresponding CrisperWhisper transcriptions. As a result, across all matching conditions, the former is evaluated on audio recordings, whereas the latter is evaluated on the corresponding CrisperWhisper transcriptions. Due to space constraints, we report only PCC; however, SRC and RMSE exhibit similar trends.
Table \ref{tab:results_baselines} reports the results on \emph{Real$_{\tt Spoken}$}, \emph{Synth$_{\tt Trs}$}, and \emph{Matched$_{\tt Spkr}$}, for which there is only one reference score label, as both text and voice originate from the same learner. The results in this table serve as best-case references with best results achieved, as expected, on \emph{Real$_{\tt Spoken}$} test data. \emph{Synth$_{\tt Trs}$} leads to a drop in performance, indicating that the TTS system already introduces some degradation even without data augmentation with different speakers. For \emph{Matched$_{\tt Spkr}$}, the drop in performance is more pronounced for ModernBERT (0.195 PCC points) than for HuBERT (0.081 PCC points). We discuss this difference in more detail below, after analysing the results reported in Table \ref{tab:results_all}.

\begin{table}[ht!]
\small
\centering
\setlength{\tabcolsep}{1.3pt} 
\begin{tabular}{l|c|c|c}
\hline
\makecell{\textbf{Linguistic}\\\textbf{content source}} 
& \textbf{Test Set} & \textbf{HuBERT} & \textbf{MBERT} \\
\hline

\multirow{2}{*}{Spoken}
& Real$_{\tt Spoken}$ & 0.677  & 0.684  \\
& Synth$_{\tt Trs}$ & 0.570  & 0.653  \\
\hline

Written
& Matched$_{\tt Spkr}$ & 0.596  & 0.489  \\
\hline

\end{tabular}

\caption{Baseline - no synthetic training data - grading results on real and matched transcript/speaker test sets in terms of PCC. One reference score label per test set.}
\label{tab:results_baselines}
\end{table}

Across all the other pairing strategies, as described in Section \ref{sec:data_aug}, each synthetic response is generated by combining a text with a voice. Therefore, two reference scores (i.e., one for audio and one for text) can be associated with a generated response. Accordingly, Table~\ref{tab:results_all} reports grading performance with respect to both labels.
As expected, performance is generally higher when HuBERT-based predictions are evaluated on audio and ModernBERT-based predictions are evaluated on text, as these represent the most appropriate input modalities for each model (see quadrants highlighted in yellow).

\begin{table}[ht!]
\small
\centering

\begin{tabular}{@{\hspace{0.1pt}}c@{\hspace{1pt}}|c|c|c}
\hline
\textbf{Model} & \textbf{Test Set} & \textbf{Audio ref.} & \textbf{Text ref.} \\
\hline

\multirow{4}{*}{HuBERT} 
& Matched$_{\tt CEFR \& L1}$ 
& \cellcolor{yellow!20} $0.544_{\pm{.02}}$ 
& $0.553_{\pm{.02}}$ \\

& Matched$_{\tt CEFR}$ 
& \cellcolor{yellow!20}
   $0.560_{\pm{.02}}$
&  $0.534_{\pm{.01}}$ \\

& Matched$_{\tt L1}$ 
& \cellcolor{yellow!20} $0.542_{\pm{.01}}$ 
& $0.399_{\pm{.05}}$ \\

& Random 
& \cellcolor{yellow!20} $0.554_{\pm{.01}}$ 
& $0.054_{\pm{.11}}$ \\

\hline\hline

\multirow{4}{*}{MBERT} 
& Matched$_{\tt CEFR \& L1}$ 
& $0.454_{\pm{.02}}$ 
& \cellcolor{yellow!20} $0.484_{\pm{.01}}$ \\

& Matched$_{\tt CEFR}$ 
&  $0.468_{\pm{.03}}$
& \cellcolor{yellow!20}
   $0.474_{\pm{.02}}$ \\

& Matched$_{\tt L1}$ 
& $0.331_{\pm{.08}}$ 
& \cellcolor{yellow!20} $0.521_{\pm{.02}}$ \\

& Random 
& $0.033_{\pm{.10}}$ 
& \cellcolor{yellow!20} $0.522_{\pm{.02}}$ \\

\hline

\end{tabular}

\caption{Baseline - no synthetic training data - PCC grading results on matched CEFR and/or L1 or random paired test sets. Two reference score labels per test set.}
\label{tab:results_all}

\end{table}

Overall, the choice of linguistic content source has relatively little effect on HuBERT, while the choice of synthetic voice has relatively little effect on ModernBERT, as reflected by the comparatively small differences across matching conditions in the yellow quadrants. This suggests that the matching strategy does not have a substantial or consistent effect on either grader when evaluated on its corresponding input modality.

For our data augmentation setting, however, the downstream grading system is not known a priori. We therefore favour matching strategies that provide stable performance across both reference types and both grading modalities. Under this criterion, \emph{Random} fails, while \emph{Matched$_{\tt CEFR\&L1}$} and \emph{Matched$_{\tt CEFR}$} show relatively consistent performance across the two reference types for both the text- and audio-based graders. \emph{Matched$_{\tt L1}$} exhibits less stable performance across evaluation settings. Since there is no substantial difference between \emph{Matched$_{\tt CEFR\&L1}$} and \emph{Matched$_{\tt CEFR}$}, we select the latter for our data augmentation experiments, as it requires less information: only the CEFR level of the source voice and the written response needs to be known.

In Section \ref{sec:exp_training}, we additionally compare the performance of systems trained using data generated under the \emph{Random} matching strategy. This provides a complementary analysis of the extent to which the choice of matching strategy affects the performance of models trained on the resulting synthetic data, rather than only their evaluation on the held-out test set.

Focusing on model-specific behaviour, a comparison of the results reported in Table~\ref{tab:results_baselines} and Table \ref{tab:results_all} shows that the HuBERT-based grader shows only a reasonable drop in PCC when evaluated on the synthetic condition \emph{Matched$_{\tt Spkr}$} compared to the \emph{Real$_{\tt Spoken}$} setting. In contrast, the ModernBERT-based grader shows a substantially larger degradation (Table~\ref{tab:results_baselines}). This degradation is not recovered by the matching strategies that we applied (Table~\ref{tab:results_all}). We hypothesise that this discrepancy arises from the linguistic gap between written and spoken language. Although the TTS system introduces some hesitations and repetitions, written responses still differ fundamentally from spoken production: they lack discourse markers, reformulations, and other indicators of online speech planning such as uncertainty or self-repair. Moreover, written responses tend to exhibit higher syntactic complexity and lexical richness, reflecting greater planning time compared to spontaneous speech.

Our hypothesis is supported by the analysis reported in Table~\ref{tab:analysis}, which presents the textual metrics described in Section \ref{sec:evaluation_metrics} for automatic transcriptions derived from real and synthetic test data. For simplicity, we report results only for one draw of the \emph{Matched$_{\tt CEFR}$} condition. Comparing \emph{Real}  and \emph{Matched$_{\tt CEFR}$} columns, transcriptions of synthetic speech derived from written responses are, on average, shorter than those derived from real learner speech (block Txt.~len.). As shown in the MDD block, they also tend to show a slightly higher syntactic complexity than real learner speech. Furthermore, they contain substantially fewer hesitations (block Hes.~rt.), reflecting the fact that TTS-generated speech does not fully capture the disfluency patterns characteristic of spontaneous learner production. At the same time, synthetic transcriptions exhibit greater lexical diversity (block MATTR) and contain a larger number of difficult words (block Dif.~wds.). This is expected, as the synthetic spoken responses are generated from written productions, which are typically more sophisticated and carefully planned than spontaneous speech. These findings support our hypothesis that the linguistic characteristics of written language are largely preserved in the synthetic responses, despite the fact that the TTS randomly introduces some hesitations and repetitions.

\begin{table*}[ht!]
\footnotesize

\centering
\setlength{\tabcolsep}{2.8pt} 
\scriptsize
\begin{tabular}{l|c|ccc|ccc|ccc|ccc|ccc}
\hline
\textbf{}
& \textbf{Count}
& \multicolumn{3}{c|}{\textbf{Txt. len.}}
& \multicolumn{3}{c|}{\textbf{MDD}}
& \multicolumn{3}{c|}{\textbf{Hes. rt. (\%)}}
& \multicolumn{3}{c|}{\textbf{MATTR}}
& \multicolumn{3}{c}{\textbf{Dif. wds.}} \\
\cline{3-17}
&
& \textbf{Real} & \textbf{Matched$_{\tt CEFR}$} & \textbf{+Spf}
& \textbf{Real} & \textbf{Matched$_{\tt CEFR}$} & \textbf{+Spf}
& \textbf{Real} & \textbf{Matched$_{\tt CEFR}$} & \textbf{+Spf}
& \textbf{Real} & \textbf{Matched$_{\tt CEFR}$} & \textbf{+Spf}
& \textbf{Real} & \textbf{Matched$_{\tt CEFR}$} & \textbf{+Spf} \\
\hline

A  &  9
   & 249.8  & 155.8  & 263.9 
   & 2.38  & 2.48  & 2.35 
   & 8.2 & 2.7  & 15.3
   & 0.56  & 0.67  & 0.61 
   & 9.0  & 11.8  & 11.0 
   \\

B1 &  15
   & 234.7  & 178.8  & 322.3 
   & 2.61  & 2.68  & 2.37 
   & 4.8  & 1.0  & 11.9 
   & 0.65 & 0.72  & 0.67 
   & 14.6  & 16.9  & 16.8 
   \\

B2 &  25
   & 284.7  & 190.2  & 330.4 
   & 2.69  & 2.62 & 2.32 
   & 5.5  & 0.8  & 10.8 
   & 0.65  & 0.72  & 0.68 
   & 15.9  & 15.6  & 16.3 
   \\

C1 &  28
   & 333.9  & 249.6  & 434.8 
   & 2.64  & 2.86  & 2.46 
   & 5.7  & 0.9  & 7.9 
   & 0.66  & 0.70  & 0.71 
   & 17.9  & 23.9  & 25.5 
   \\

C2 &  23
   & 328.9  & 222.4  & 443.9 
   & 2.56 & 2.68  & 2.40 
   & 3.9  & 0.7  & 7.8 
   & 0.68  & 0.73  & 0.72 
   & 19.0  & 22.3  & 26.3 
   \\

\hline
\end{tabular}

\caption{Text statistics across CEFR levels for CrisperWhisper transcriptions from \emph{Real$_{\tt Spoken}$}, \emph{Matched$_{\tt CEFR}$}, and \emph{Matched$_{\tt CEFR}$+Speechify} (+Spf) data. Abbreviations are explained in Section \ref{sec:evaluation_metrics}.}
\label{tab:analysis}

\end{table*}


\subsection{Effects of speechification}
\label{sec:exp_speechify}

To address the gap between written and spoken language, we revisit the written responses and apply the speechification procedure described in Section~\ref{sec:speechify}, transforming them into spoken-style learner transcripts using GPT-4.1 prior to speech synthesis. To assess whether semantic content is preserved after speechification, as described in Section \ref{sec:evaluation_metrics}, we compute the cosine similarity between sentence embeddings extracted from the original (\emph{Real$_{\tt Written}$}) and speechified (\emph{Real$_{\tt Written}$+Speechify}) written responses. 
The value of $0.853_{\pm{.052}}$ indicates that the speechified responses remain highly similar to the original texts, suggesting that the speechification process largely preserves semantic content. For comparison, the cosine similarity between real written and spoken texts from the same learners is $0.813_{\pm{.080}}$.

In addition to semantic similarity, we analyse the \emph{+Spf} column in Table~\ref{tab:analysis}, which reports the text statistics described in Section~\ref{sec:txt_anl} for speechified \emph{Matched$_{\tt CEFR}$} transcriptions. Overall, speechification shifts several properties of the generated transcripts towards those observed in real learner speech. For MDD, the speechified texts show lower values than the real spoken texts across all CEFR levels, suggesting that the procedure may over-simplify syntactic structure, while still moving away from the higher MDD values observed in the non-speechified transcriptions. Similarly, the hesitation rate increases substantially after speechification and exceeds the values observed in real speech, indicating that the model may overproduce hesitation phenomena. In contrast, MATTR generally moves towards the values observed in real speech. Speechification also substantially increases text length relative to the non-speechified \emph{Matched$_{\tt CEFR}$} transcripts, bringing it closer to, and at several levels beyond, the length of real spoken responses. Finally, the number of difficult words remains relatively stable across conditions, suggesting that speechification primarily modifies the form of the responses rather than substantially changing their lexical difficulty. Together with the high semantic similarity reported above, these results indicate that speechification introduces several characteristics of spoken language while preserving the core content of the original responses, although some features, particularly hesitations and syntactic complexity, appear to be exaggerated.

Table~\ref{tab:results_aug} compares the performance obtained on \emph{Matched$_{\tt CEFR}$} data before and after speechification. As illustrated in Section \ref{sec:speechify}, speechification is conditioned on the CEFR level associated with the learner's text. Importantly, this CEFR information is used here only as a control variable for the generation process and for analysing whether proficiency-specific speech characteristics can be reproduced; it is not provided to the grading models and is not used to determine their predictions. Thus, the use of CEFR conditioning in this analysis should not be interpreted as providing additional proficiency information to the downstream grading systems. We also report the results when running the speechification process without injecting the CEFR-related labels into the prompt given to the LLM.
The results indicate that CEFR-conditioned speechification substantially improves the quality of the synthetic data, yielding large gains for the ModernBERT-based grader and more moderate, yet consistent, improvements for the HuBERT-based system.
These findings suggest that modelling the discourse- and fluency-related characteristics is at least as important as preserving acoustic characteristics when generating synthetic data for speaking assessment. The fact that, for ModernBERT, the results for \emph{Matched$_{\tt CEFR}$+Speechify} are even higher than those for \emph{Real} (see Table \ref{tab:results_baselines}) should not be surprising, as the two settings are not directly comparable: the former, despite speechification, is generated starting from written productions, whereas the latter consists of authentic spoken responses. Therefore, the linguistic content of the two sets is different.

The analysis in this section is intended to characterise the effect of speechification and its conditioning, rather than to establish its effectiveness as a training data augmentation strategy. We therefore provide a separate ablation experiment in Table \ref{tab:results_sp_vs_nosp} at the end of Section~\ref{sec:exp_training}, where HuBERT and ModernBERT are trained with and without speechified synthetic data and evaluated on the real test set. This experiment provides a direct assessment of the contribution of speechification to grading performance under the final training and evaluation setting.

\begin{table}[ht!]

\setlength{\tabcolsep}{2.8pt} 
\footnotesize
\centering
\begin{tabular}{c|c|c|c}
\hline
\textbf{Test Set} & \textbf{CEFR in Spf} & \textbf{HuBERT} & \textbf{MBERT} \\
\hline

 Matched$_{\tt CEFR}$ & - & $0.560_{\pm{.02}}$  & $0.474_{\pm{.04}}$        \\
 Matched$_{\tt CEFR}$+Spf & $\times$ & $0.542_{\pm{.02}}$   & $0.457_{\pm{.01}}$       \\
 Matched$_{\tt CEFR}$+Spf & $\checkmark$ & \bm{$0.583_{\pm{.01}}$}   & \bm{$0.715_{\pm{.01}}$}       \\
 
\hline
\end{tabular}
\caption{Analysis on synthetic test sets in terms of PCC: \emph{Matched$_{\tt CEFR}$} vs \emph{Matched$_{\tt CEFR}$+Speechify} (with and without CEFR-conditioning).}
\label{tab:results_aug}

\end{table}

\begin{figure}[ht]
    \centering

    \begin{subfigure}{}
        \centering
        \includegraphics[width=0.7\linewidth]{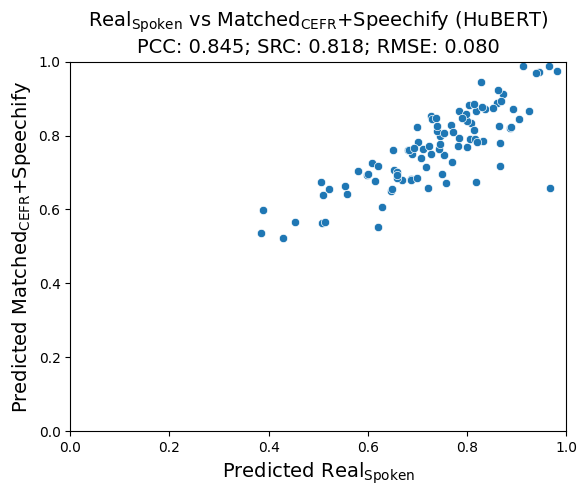}
    \end{subfigure}
    
    \begin{subfigure}{}
        \centering
        \includegraphics[width=0.7\linewidth]{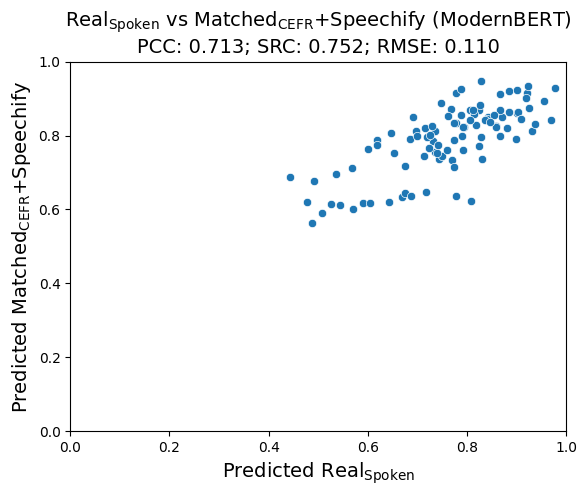}
    \end{subfigure}
    
    \caption{\emph{Real$_{\tt Spoken}$} vs \emph{Matched$_{\tt CEFR}$+Speechify} predictions.}
    \label{fig:scatterplots}
\end{figure}

Additionally, in Figure~\ref{fig:scatterplots}, we report scatterplots between predictions on \emph{Real$_{\tt Spoken}$} and predictions on \emph{Matched$_{\tt CEFR}$+Speechify} for both graders, together with the corresponding PCC, SRC, and RMSE values. The higher performance observed for the HuBERT-based system can be explained as follows: HuBERT operates directly on acoustic signals, which are largely preserved through the TTS process, as shown in Table \ref{tab:results_WER} (see Cosine Similarity column), despite differences in linguistic content. In contrast, ModernBERT operates on text and is therefore more sensitive to differences in linguistic realisation, since \emph{Real$_{\tt Spoken}$} derives from spoken responses, while \emph{Matched$_{\tt CEFR}$+Speechify}, despite speechification, still originates from a written source.

  \subsection{Training and testing with augmented data}
  \label{sec:exp_training}
At this point, we can use the augmented training data to train our grading systems and evaluate them on the real test set.
 In Table~\ref{tab:results_final}, we compare different training configurations. \emph{Real$_{\tt Spoken}$} refers to models trained exclusively on the real training spoken data. \emph{Syn} denotes training on synthetic data generated using the \emph{Matched$_{\tt CEFR}$+Speechify} pipeline by matching the 410 voices with the 383 written responses in the training set (see Table \ref{tab:data}). For the audio-based grader, this results in 410 synthetic speech samples, with 27 (i.e., $410 - 383$) written responses reused across different voices. Conversely, for the text-based grader, training is performed on the automatic transcriptions obtained after synthesising the 383 unique written responses without duplicates. Finally, \emph{Syn$_{\tt Extra}$} extends \emph{Syn} by additionally incorporating the 2,820 written responses without spoken counterparts (i.e., \emph{Extra} in Table \ref{tab:data}).
 
Table~\ref{tab:results_final} reports the results on the \emph{Real$_{\tt Spoken}$} test set. As can be observed, data augmentation leads to consistent performance improvements across all evaluation metrics for both models. We hypothesise that this gain is particularly pronounced for the ModernBERT-based system due to the substantially increased amount of training text available. In the \emph{Real$_{\tt Spoken}$+Syn$_{\tt Extra}$} setting, the model effectively benefits from a total of 3,613 training automatic transcriptions (i.e., 410 from \emph{Real$_{\tt Spoken}$}, 383 from \emph{Syn}, and the remaining 2,820 in \emph{Syn$_{\tt Extra}$}). In contrast, the HuBERT-based system, although exposed to the same augmented synthetic speech pipeline, remains constrained by the diversity of available speaker identities. Specifically, it still relies on the 410 unique voices present in the spoken training set, which limits the effective increase in variability compared to the text-based model. However, for both models, the \emph{Real$_{\tt Spoken}$}+\emph{Syn}$_{\tt Extra}$ setting significantly outperforms the \emph{Real$_{\tt Spoken}$} baseline.\footnote{We used a paired permutation test with 10,000 within-pair swaps comparing \emph{Real$_{\tt Spoken}$} to \emph{Real$_{\tt Spoken}$}+\emph{Syn}$_{\tt Extra}$. This yielded $p = 0.041$ for HuBERT and $p = 0.011$ for ModernBERT.}

 \begin{table}[ht!]
 \small
 
\centering
\setlength{\tabcolsep}{4.4pt} 
\begin{tabular}{c|c|c|c|c}
\hline
\textbf{Model} & \textbf{Training Set} & \textbf{PCC} & \textbf{SRC} & \textbf{RMSE} \\
\hline
\multirow{4}{*}{HuBERT} 
& Real$_{\tt Spoken}$ & 0.677   & 0.634   & 0.133   \\
\cline{2-5}

& Syn & 0.663  & 0.629    & 0.133   \\

& Syn$_{\tt Extra}$  & 0.687   & 0.656    & 0.134   \\

& Real$_{\tt Spoken}$+Syn$_{\tt Extra}$ & \textbf{0.739}  & \textbf{0.692}   & \textbf{0.119}  \\

\hline
\multirow{4}{*}{MBERT} 
& Real$_{\tt Spoken}$ & 0.684  & 0.656   & 0.132   \\
\cline{2-5}

& Syn & 0.643  & 0.577     & 0.136   \\

& Syn$_{\tt Extra}$ & 0.663   & 0.607   & 0.132   \\

& Real$_{\tt Spoken}$+Syn$_{\tt Extra}$ & \textbf{0.769}  & \textbf{0.710}  & \textbf{0.113} \\

\hline
\multirow{4}{*}{Both} 
& Real$_{\tt Spoken}$ & 0.730   & 0.699    & 0.121    \\
\cline{2-5}

& Syn & 0.696   & 0.652      & 0.126    \\

& Syn$_{\tt Extra}$ & 0.717    & 0.686    & 0.123    \\

& Real$_{\tt Spoken}$+Syn$_{\tt Extra}$ & \textbf{0.795}  & \textbf{0.775}  & \textbf{0.107} \\

\hline

\end{tabular}
\caption{Results on real test data using different training sets. \emph{Syn} and \emph{Syn$_{\tt Extra}$} use the \emph{Matched$_{\tt CEFR}$+Speechify} process.}
\label{tab:results_final}
\end{table}

Interestingly, the combination of HuBERT and ModernBERT\footnote{I.e., the average of the two grading systems predictions.} using \emph{Real$_{\tt Spoken}$}+\emph{Syn}$_{\tt Extra}$ achieves the best performance showing that the two models carry complementary information. As we observed for the two models taken individually, the \emph{Real$_{\tt Spoken}$}+\emph{Syn}$_{\tt Extra}$ setting is significantly better than the \emph{Real$_{\tt Spoken}$} baseline.\footnote{The paired permutation test with 10,000 within-pair swaps yielded $p=0.009$.}

One of the leitmotifs of this paper is that the downstream grading system is not known a priori. However, if the target system is known to be text-based, TTS is not necessarily required. To examine this, we train a ModernBERT-based grader directly on the original \emph{Real$_{\tt Spoken}$} CrisperWhisper transcriptions together with the \emph{Syn$_{\tt Extra}$} data before the synthesis step, i.e., using the speechified written texts directly. This yields PCC = 0.778, SRC = 0.732, and RMSE = 0.110, compared with PCC = 0.769, SRC = 0.710, and RMSE = 0.113 when the same synthetic texts are first converted to speech and subsequently transcribed. The slightly better performance obtained without TTS suggests that speech synthesis and subsequent ASR introduce some noise that is detrimental to text-based grading. Thus, while TTS is necessary when augmenting an acoustic grading system, it appears to be an unnecessary intermediate step for a text-based grader when the latter can operate directly on the speechified transcripts.

 \begin{table}[ht!]
 \small
 
\centering
\setlength{\tabcolsep}{4.8pt} 
\begin{tabular}{c|c|c|c|c}
\hline
\textbf{Model} & \textbf{Matching} & \textbf{PCC} & \textbf{SRC} & \textbf{RMSE} \\
\hline
\multirow{3}{*}{HuBERT}

& Matched$_{ \tt CEFR}$ & \textbf{0.663}  & \textbf{0.629}    & \textbf{0.133}   \\

& Random$_{\tt Audio\_Label}$ & 0.632   & 0.614     & 0.139    \\ 
& Random$_{\tt Text\_Label}$ & 0.322   & 0.342    & 0.175   \\ 
\hline
\multirow{3}{*}{MBERT}

& Matched$_{ \tt CEFR}$ & \textbf{0.643}  & 0.577     & \textbf{0.136}   \\

& Random$_{\tt Audio\_Label}$ & 0.164  & 0.061       & 0.185    \\

& Random$_{\tt Text\_Label}$ & 0.632   & \textbf{0.586}       & 0.140   \\


\hline

\hline

\end{tabular}
\caption{Results on real test data comparing \emph{Matched$_{\tt CEFR}$+Speechify} to \emph{Random+Speechify} training data using the \emph{Syn} training partition.}
\label{tab:results_random_vs_cefr}
\end{table}

Furthermore, as mentioned in Section \ref{sec:exp_test}, we consider a different matching strategy, i.e., \emph{Random+Speechify}, to generate synthetic training data. Under this matching strategy, the TTS-generated responses might have two very different text and speaker labels. We therefore train separate graders on the audio (\emph{Random}$_{\tt Audio\_Label}$) and the text labels (\emph{Random}$_{\tt Text\_Label}$) using the \emph{Syn} training data partition. We compare this matching strategy to \emph{Matched$_{\tt CEFR}$} in Table \ref{tab:results_random_vs_cefr}. These results suggest that matching synthetic voices and texts by CEFR provides a more reliable alignment between the augmented samples and the proficiency labels than random matching. \emph{Matched$_{\tt CEFR}$} achieves the best PCC and RMSE for both models and the best SRC for HuBERT, with only a small exception for ModernBERT SRC, where \emph{Random$_{\tt TextLabel}$} performs marginally better. More importantly, CEFR-based matching avoids a fundamental ambiguity associated with random matching: when the voice and text originate from learners with different proficiency levels, it is unclear whether the proficiency label of the synthetic sample should be determined by the voice or by the text. This is particularly problematic because the appropriate choice depends on the modality used by the downstream grader. By matching the voice and text at the same CEFR level (see Section \ref{sec:grading_sys}), \emph{Matched$_{\tt CEFR}$} ensures that both modalities are associated with the same proficiency level, removing the need to make this decision and making the resulting synthetic data independent of the modality of the downstream grader. This is exactly where \emph{Random} is shown to fail (see row \emph{Random}$_{\tt Text\_Label}$ for HuBERT and row \emph{Random}$_{\tt Audio\_Label}$ for ModernBERT).

Finally, as mentioned in Section \ref{sec:exp_speechify}, we conduct an ablation study in order to provide a direct assessment of the contribution of speechification to grading performance. Specifically, we consider the \emph{Syn} training partition using the \emph{Matched$_{\tt CEFR}$} setting with and without the speechification process. As can be observed in Table \ref{tab:results_sp_vs_nosp}, speechification consistently improves performance across all three evaluation metrics for both grading models. The improvement is substantially larger for ModernBERT, suggesting that the spoken-language phenomena introduced through speechification are particularly beneficial for the text-based grading system. This finding supports the analysis presented in Table \ref{tab:results_aug} in Section~\ref{sec:exp_speechify}, while providing direct evidence that speechification contributes to downstream grading performance when used as a training data augmentation strategy.

 \begin{table}[ht!]
 \small
 
\centering
\setlength{\tabcolsep}{4.8pt} 
\begin{tabular}{c|c|c|c|c}
\hline
\textbf{Model} & \textbf{Speechification} & \textbf{PCC} & \textbf{SRC} & \textbf{RMSE} \\
\hline
\multirow{2}{*}{HuBERT}

& $\checkmark$ & \textbf{0.663}  & \textbf{0.629}    & \textbf{0.133}   \\

& $\times$ & 0.640    & 0.621      & 0.139    \\ 
 
\hline
\multirow{2}{*}{MBERT}

& $\checkmark$ & \textbf{0.643}  & \textbf{0.577}     & \textbf{0.136}   \\

& $\times$ & 0.546   & 0.522        & 0.148     \\


\hline

\hline

\end{tabular}
\caption{Results on real test data comparing \emph{Matched$_{\tt CEFR}$} training data with and without speechification using the \emph{Syn} training partition.}
\label{tab:results_sp_vs_nosp}
\end{table}

\section{Conclusions}

In this work, we investigated how to generate synthetic L2 speech for proficiency assessment. We proposed a cross-modal augmentation framework that speechifies written L2 responses using an LLM followed by TTS synthesis. Using HuBERT- and ModernBERT-based graders, we showed that pairing speakers and texts by proficiency level is a principled and effective strategy for data augmentation. We further found that raw written text creates a strong mismatch with spoken language, while speechification substantially reduces this gap and improves performance. Overall, synthetic data consistently improve both audio- and text-based grading systems, highlighting the importance of modelling spoken linguistic structure in addition to speaker characteristics.

Future work will explore LLM-generated responses directly, removing reliance on written sources, following \citet{wang2025novel} and \citet{voskoboinik2025enhancing}.

\section*{Limitations}

Our experiments are based on COREFL and therefore involve a relatively limited number of speakers and approximately 16.7 hours of spoken training data. The extent to which our findings generalise to other L2 populations, proficiency distributions, languages, or assessment settings remains to be established. In particular, our experiments use a specific combination of LLM, TTS, ASR, and proficiency models, and different model architectures may respond differently to synthetic data.

A further limitation concerns the calibration of the speechification process. Although the LLM is instructed to modulate disfluency rates and syntactic complexity according to the learner's CEFR level, these instructions are qualitative rather than derived from the empirical distributions observed in real learner speech. As shown in Table \ref{tab:analysis}, the resulting speechified transcriptions overestimate hesitation rates relative to authentic spoken productions at every proficiency level and, conversely, tend to underestimate syntactic complexity as measured by MDD. Explicitly calibrating the speechification prompt through few-shot prompting or against corpus-derived, level-specific statistics (e.g., target hesitation rates, MDD ranges) is a promising direction for future work, but is not explored in this study.

\section*{Ethics Statement}

The use of voice cloning to generate synthetic L2 speech introduces ethical and privacy risks that require particular consideration. A cloned or synthesised voice may closely resemble that of a real speaker, raising concerns related to consent, speaker identity, impersonation, and unauthorised reuse of voice data.

In this work, synthetic speech is generated solely for research and assessment-data augmentation. Voice cloning is not used to provide learner feedback, for example, by generating a synthesised version of a learner's utterance with corrected pronunciation or grammatical errors. Instead, its sole purpose is to increase the amount and diversity of training data available to the grading systems.

\section*{Acknowledgments}

This paper reports on research supported by Cambridge University Press \& Assessment, a department of The Chancellor, Masters, and Scholars of the University of Cambridge. The authors would like to thank the ALTA Spoken Language Processing Technology Project Team for general discussions and contributions to the evaluation infrastructure.

Generative AI tools, namely ChatGPT and Gemini, were used only to review grammar and improve the fluency of the text.

\bibliographystyle{acl_natbib}
\bibliography{nlp4call}

@article{crossley2023english,
  title={{The English Language Learner Insight, Proficiency and Skills Evaluation (ELLIPSE) Corpus}},
  author={Crossley, Scott and Tian, Yu and Baffour, Perpetual and Franklin, Alex and Kim, Youngmeen and Morris, Wesley and Benner, Meg and Picou, Aigner and Boser, Ulrich},
  journal={International Journal of Learner Corpus Research},
  volume={9},
  number={2},
  pages={248--269},
  year={2023},
  publisher={John Benjamins Publishing Company Amsterdam/Philadelphia}
}

@article{hsu2021hubert,
  title={Hubert: Self-supervised speech representation learning by masked prediction of hidden units},
  author={Hsu, Wei-Ning and Bolte, Benjamin and Tsai, Yao-Hung Hubert and Lakhotia, Kushal and Salakhutdinov, Ruslan and Mohamed, Abdelrahman},
  journal={IEEE/ACM Transactions on Audio, Speech, and Language Processing},
  volume={29},
  pages={3451--3460},
  year={2021},
  publisher={IEEE}
}

@inproceedings{wang2021,
  author={Wang, X. and Evanini, K. and Qian, Y. and Mulholland, M.},
  booktitle={Proc. 2021 IEEE Spoken Language Technology Workshop (SLT)}, 
  title={Automated Scoring of Spontaneous Speech from Young Learners of English Using Transformers}, 
  year={2021},
  volume={},
  number={},
  pages={705-712}}

@inproceedings{raina2020universal,
  author={Vyas Raina and Mark J.F. Gales and Kate M. Knill},
  title={{Universal Adversarial Attacks on Spoken Language Assessment Systems}},
  year=2020,
  booktitle={Proc. Interspeech 2020},
  pages={3855--3859},
  doi={10.21437/Interspeech.2020-1890}
}

@article{blanchard2013toefl11,
  title={{TOEFL11: A corpus of non-native English}},
  author={Blanchard, Daniel and Tetreault, Joel and Higgins, Derrick and Cahill, Aoife and Chodorow, Martin},
  journal={ETS Research Report Series},
  volume={2013},
  number={2},
  pages={i--15},
  year={2013},
  publisher={Wiley Online Library}
}

@article{voskoboinik2025enhancing,
  title={{Enhancing second language speaking assessment: Integrating large language models for Finnish and Finland Swedish proficiency scoring}},
  author={Voskoboinik, Ekaterina and von Zansen, Anna and Phan, Nhan Chi and Getman, Yaroslav and Gr{\'o}sz, Tam{\'a}s and Kurimo, Mikko},
  journal={Language Testing},
  volume={42},
  number={4},
  pages={508--538},
  year={2025},
  publisher={SAGE Publications Sage UK: London, England}
}

@inproceedings{wang2025novel,
  author={Wang, Chung-Chun and Lin, Jhen-Ke and Lu, Hao-Chien and Lin, Hong-Yun and Chen, Berlin},
  title={A Novel Data Augmentation Approach for Automatic Speaking Assessment on Opinion Expressions},
  booktitle={Proc. 10th Workshop on Speech and Language Technology in Education (SLaTE)},
  pages={199-203},
  year={2025}
}

@article{yan2025disfluency,
  title={{Disfluency doesn’t happen in isolation: Exploring how individual disfluency features co-occur in L2 speaking performances}},
  author={Yan, Xun and Chuang, Ping-Lin and Pan, Yulin and Cai, Huiying and Staples, Shelley and Bertho, Mariana Centanin},
  journal={Studies in Second Language Acquisition},
  volume={47},
  number={2},
  pages={560--591},
  year={2025},
  publisher={Cambridge University Press}
}

@article{huang2023development,
  title={{Development of the use of discourse markers across different fluency levels of CEFR: A learner corpus analysis}},
  author={Huang, Lan-fen and Lin, Yen-liang and Gr{\'a}f, Tom{\'a}{\v{s}}},
  journal={Pragmatics},
  volume={33},
  number={1},
  pages={49--77},
  year={2023},
  publisher={International Pragmatics Association (IPrA)}
}

@inproceedings{peng21,
  author={L. Peng and K. Fu and B. Lin and D. Ke and J. Zhan},
  title={{A Study on Fine-Tuning wav2vec2.0 Model for the Task of Mispronunciation Detection and Diagnosis}},
  year=2021,
  booktitle={Proc. Interspeech 2021},
  pages={4448--4452},
  doi={10.21437/Interspeech.2021-1344}
}

@inproceedings{xu21,
  author={X. Xu and Y. Kang and S. Cao and B. Lin and L. Ma},
  title={{Explore wav2vec 2.0 for Mispronunciation Detection}},
  year=2021,
  booktitle={Proc. Interspeech 2021},
  pages={4428--4432},
  doi={10.21437/Interspeech.2021-777}
}

@inproceedings{eesung,
  title     = {{Automatic Pronunciation Assessment using Self-Supervised Speech Representation Learning}},
  author    = {Eesung Kim and Jae-Jin Jeon and Hyeji Seo and Hoon Kim},
  year      = {2022},
  booktitle = {{Interspeech 2022}},
  pages     = {1411--1415},
  doi       = {10.21437/Interspeech.2022-10245},
  issn      = {2958-1796},
}

@incollection{ishikawa2011,
title={ A New horizon in learner corpus studies: The aim of the {ICNALE} Project},
author={Ishikawa, S.},
booktitle={Corpora and language technologies in teaching, learning and research},
publisher={University of Strathclyde Press},
place = {Glasgow},
year={2011},
pages={3-11}
}

@book{cefr2001,
author= {{Council of Europe}},
title={Common European Framework of Reference for Languages: Learning, Teaching, Assessment},
address={Cambridge},
publisher={Cambridge University Press},
year=2001,
url={https://rm.coe.int/1680459f97}
}

@inproceedings{devlin2018,
    title = "{BERT}: Pre-training of Deep Bidirectional Transformers for Language Understanding",
    author = "Devlin, Jacob  and
      Chang, Ming-Wei  and
      Lee, Kenton  and
      Toutanova, Kristina",
    booktitle = "Proc. 2019 Conference of the North {A}merican Chapter of the Association for Computational Linguistics: Human Language Technologies, Volume 1 (Long and Short Papers)",
    month = jun,
    year = "2019",
    address = "Minneapolis, Minnesota",
    url = "https://aclanthology.org/N19-1423",
    doi = "10.18653/v1/N19-1423",
    pages = "4171--4186",
}

@misc{openai2023gpt4,
      title={{GPT-4 Technical Report}}, 
      author={OpenAI},
      year={2023},
      eprint={2303.08774},
      archivePrefix={arXiv},
      primaryClass={cs.CL},
      doi={10.48550/arXiv.2303.08774}
}

@inproceedings{baevski,
  title={wav2vec 2.0: A framework for self-supervised learning of speech representations},
  year=2020,
  booktitle={NeurIPS 2020},
  pages={1--12},
  author={A. Baevski and H. Zhou and A. Mohamed and M. Auli},
}

@inproceedings{craighead2020,
    title = "Investigating the effect of auxiliary objectives for the automated grading of learner {E}nglish speech transcriptions",
    author = "Craighead, Hannah  and
      Caines, Andrew  and
      Buttery, Paula  and
      Yannakoudakis, Helen",
    booktitle = "Proc. 58th Annual Meeting of the Association for Computational Linguistics",
    month = jul,
    year = "2020",
    address = "Online",
    url = "https://aclanthology.org/2020.acl-main.206/",
    doi = "10.18653/v1/2020.acl-main.206",
    pages = "2258--2269",
    }

@inproceedings{geertzen2013automatic,
  title={Automatic linguistic annotation of large scale {L2} databases: {The EF-Cambridge Open Language Database (EFCAMDAT)}},
  author={Geertzen, Jeroen and Alexopoulou, Theodora and Korhonen, Anna},
  booktitle={Proc. 31st Second Language Research Forum},
  pages={240--254},
  year={2013},
  address={Somerville},
  url={http://www.lingref.com/cpp/slrf/2012/paper3100.pdf}
}

@inproceedings{zhao2018l2,
  title={{L2-ARCTIC}: A non-native {English} speech corpus.},
  author={Zhao, Guanlong and Sonsaat, Sinem and Silpachai, Alif and Lucic, Ivana and Chukharev-Hudilainen, Evgeny and Levis, John and Gutierrez-Osuna, Ricardo},
  booktitle={Proc. Interspeech 2018},
  pages={2783--2787},
  year={2018}
}

@inproceedings{speechocean762,
  title     = {{speechocean762: An Open-Source Non-Native English Speech Corpus for Pronunciation Assessment}},
  author    = {Junbo Zhang and Zhiwen Zhang and Yongqing Wang and Zhiyong Yan and Qiong Song and Yukai Huang and Ke Li and Daniel Povey and Yujun Wang},
  year      = {2021},
  booktitle = {{Proc. Interspeech 2021}},
  pages     = {3710--3714},
  doi       = {10.21437/Interspeech.2021-1259},
  issn      = {2958-1796},
}

@inproceedings{banno23_slate,
  author={Stefano Bannò and Katherine M Knill and Marco Matassoni and Vyas Raina and Mark Gales},
  title={{Assessment of L2 Oral Proficiency Using Self-Supervised Speech Representation Learning}},
  year=2023,
  booktitle={Proc. 9th Workshop on Speech and Language Technology in Education (SLaTE)},
  pages={126--130},
  doi={10.21437/SLaTE.2023-24}
}

@article{liu2008dependency,
  title={Dependency distance as a metric of language comprehension difficulty},
  author={Liu, Haitao},
  journal={Journal of Cognitive Science},
  volume={9},
  number={2},
  pages={159--191},
  year={2008}
}

@inproceedings{baba2024utmosv2,
  title     = {The T05 System for The {V}oice{MOS} {C}hallenge 2024: Transfer Learning from Deep Image Classifier to Naturalness {MOS} Prediction of High-Quality Synthetic Speech},
  author    = {Baba, Kaito and Nakata, Wataru and Saito, Yuki and Saruwatari, Hiroshi},
  booktitle = {IEEE Spoken Language Technology Workshop (SLT)},
  year      = {2024},
  pages     = {818--824},
  doi       = {10.1109/SLT61566.2024.10832315},
}

@article{sicorpus25,
  author = {Kate Knill and Diane Nicholls and Mark J.F. Gales and Mengjie Qian and Pawel Stroinski},
  year = {2025},
  title = {{The Speak \& Improve Corpus 2025: an L2 English Speech Corpus for Language Assessment and Feedback}},
  publisher = {Cambridge University Press & Assessment},
  url = {https://doi.org/10.17863/CAM.114333}
}

@inproceedings{qian25_challenge,
  title     = {{Speak \& Improve Challenge 2025}},
  author    = {Mengjie Qian and Kate Knill and Stefano Bannò and Siyuan Tang and Penny Karanasou and Mark Gales and Diane Nicholls},
  year      = {2025},
  booktitle = {{Proc. 10th Workshop on Speech and Language Technology in Education (SLaTE)}},
  pages     = {41--45},
  doi       = {10.21437/SLaTE.2025-9},
}

@article{nicholls2024write,
  author = {Diane Nicholls and Andrew Caines and Paula Buttery},
  year = {2024},
  title = {The {W}rite \& {I}mprove {C}orpus 2024: Error-annotated and {CEFR}-labelled essays by learners of {E}nglish},
  publisher = {Cambridge University Press & Assessment},
  url = {https://doi.org/10.17863/CAM.112997}
}

@inproceedings{warner-etal-2025-smarter,
    title = "Smarter, Better, Faster, Longer: A Modern Bidirectional Encoder for Fast, Memory Efficient, and Long Context Finetuning and Inference",
    author = {Warner, Benjamin  and
      Chaffin, Antoine  and
      Clavi{\'e}, Benjamin  and
      Weller, Orion  and
      Hallstr{\"o}m, Oskar  and
      Taghadouini, Said  and
      Gallagher, Alexis  and
      Biswas, Raja  and
      Ladhak, Faisal  and
      Aarsen, Tom  and
      Adams, Griffin Thomas  and
      Howard, Jeremy  and
      Poli, Iacopo},
    editor = "Che, Wanxiang  and
      Nabende, Joyce  and
      Shutova, Ekaterina  and
      Pilehvar, Mohammad Taher",
    booktitle = "Proc. 63rd Annual Meeting of the Association for Computational Linguistics (Volume 1: Long Papers)",
    month = jul,
    year = "2025",
    address = "Vienna, Austria",
    publisher = "Association for Computational Linguistics",
    url = "https://aclanthology.org/2025.acl-long.127/",
    doi = "10.18653/v1/2025.acl-long.127",
    pages = "2526--2547",
    ISBN = "979-8-89176-251-0"
}

@inproceedings{song2020mpnet,
  title={{MPNet: Masked and permuted pre-training for language understanding}},
  author={Song, Kaitao and Tan, Xu and Qin, Tao and Lu, Jianfeng and Liu, Tie-Yan},
  booktitle={Proc. Advances in Neural Information Processing Systems (NeuRIPS)},
  volume={33},
  pages={16857--16867},
  year={2020}
}

@inproceedings{reimers-gurevych-2019-sentence,
    title = "Sentence-{BERT}: Sentence Embeddings using {S}iamese {BERT}-Networks",
    author = "Reimers, Nils  and
      Gurevych, Iryna",
    booktitle = "Proc. 2019 Conference on Empirical Methods in Natural Language Processing and the 9th International Joint Conference on Natural Language Processing (EMNLP-IJCNLP)",
    month = nov,
    year = "2019",
    address = "Hong Kong, China",
    url = "https://aclanthology.org/D19-1410/",
    doi = "10.18653/v1/D19-1410",
    pages = "3982--3992"
}

@inproceedings{zusag24_interspeech,
  title     = {{CrisperWhisper: Accurate Timestamps on Verbatim Speech Transcriptions}},
  author    = {Mario Zusag and Laurin Wagner and Bernhad Thallinger},
  year      = {2024},
  booktitle = {{Proc. Interspeech 2024}},
  pages     = {1265--1269},
  doi       = {10.21437/Interspeech.2024-731},
  issn      = {2958-1796},
}

@misc{zhu2026omnivoiceomnilingualzeroshottexttospeech,
      title={{OmniVoice: Towards Omnilingual Zero-Shot Text-to-Speech with Diffusion Language Models}}, 
      author={Han Zhu and Lingxuan Ye and Wei Kang and Zengwei Yao and Liyong Guo and Fangjun Kuang and Zhifeng Han and Weiji Zhuang and Long Lin and Daniel Povey},
      year={2026},
      eprint={2604.00688},
      archivePrefix={arXiv},
      primaryClass={cs.CL},
      url={https://arxiv.org/abs/2604.00688}, 
}

@inproceedings{mcknight23_slate,
  title     = {{Automatic Assessment of Conversational Speaking Tests}},
  author    = {Simon W McKnight and Arda Civelekoglu and Mark Gales and Stefano Bannò and Adian Liusie and Katherine M Knill},
  year      = {2023},
  booktitle = {{Proc. 9th Workshop on Speech and Language Technology in Education (SLaTE)}},
  pages     = {99--103},
  doi       = {10.21437/SLaTE.2023-19},
  issn      = {2311-4975},
}

@inproceedings{karanasou25_slate,
  title     = {{Data Augmentation for Spoken Grammatical Error Correction}},
  author    = {Penny Karanasou and Mengjie Qian and Stefano Bannò and Mark J.F. Gales and Kate M. Knill},
  year      = {2025},
  booktitle = {{Proc. 10th Workshop on Speech and Language Technology in Education (SLaTE)}},
  pages     = {194--198},
  doi       = {10.21437/SLaTE.2025-39},
  issn      = {2311-4975},
}

@incollection{lozano2020designing,
  title={{Designing and compiling a learner corpus of written and spoken narratives: COREFL}},
  author={Lozano, Crist{\'o}bal and D{\'\i}az-Negrillo, Ana and Callies, Marcus},
  booktitle={What's in a Narrative? Variation in Storytelling at the Interface Between Language and Literacy},
      publisher = "Peter Lang Verlag",
    address = "Berlin, Germany",
    doi = "10.3726/978-3-653-05182-7",
    url = "https://www.peterlang.com/document/1049094",
    editor = "Christiane {Bongartz} and Jacopo {Torregrossa}",
  pages={21--46},
  year={2021}
}

@inproceedings{Bredin2020,
  Title = {{pyannote.audio: neural building blocks for speaker diarization}},
  Author = {{Bredin}, Herv{\'e} and {Yin}, Ruiqing and {Coria}, Juan Manuel and {Gelly}, Gregory and {Korshunov}, Pavel and {Lavechin}, Marvin and {Fustes}, Diego and {Titeux}, Hadrien and {Bouaziz}, Wassim and {Gill}, Marie-Philippe},
  Booktitle = {Proc. ICASSP 2020, IEEE International Conference on Acoustics, Speech, and Signal Processing},
  Address = {Barcelona, Spain},
  Month = {May},
  Year = {2020},
}

@article{Covington2010,
author = {Michael A. Covington and Joe D. McFall},
title = {{Cutting the Gordian Knot: The Moving-Average Type–Token Ratio (MATTR)}},
journal = {Journal of Quantitative Linguistics},
volume = {17},
number = {2},
pages = {94--100},
year = {2010},
publisher = {Routledge},
doi = {10.1080/09296171003643098},


URL = { 
    
        https://doi.org/10.1080/09296171003643098
    
    

},
eprint = { 
    
        https://doi.org/10.1080/09296171003643098
    
    

}

}

@book{zechner2019automated,
  title={{Automated speaking assessment: Using language technologies to score spontaneous speech}},
  author={Zechner, Klaus and Evanini, Keelan},
  year={2019},
  publisher={Routledge}
}

@article{biber2002speaking,
  title={Speaking and writing in the university: A multidimensional comparison},
  author={Biber, Douglas and Conrad, Susan and Reppen, Randi and Byrd, Pat and Helt, Marie},
  journal={TESOL Quarterly},
  volume={36},
  number={1},
  pages={9--48},
  year={2002},
  publisher={Wiley Online Library}
}

@inproceedings{coulange-etal-2024-corpus,
    title = "A Corpus of Spontaneous {L}2 {E}nglish Speech for Real-situation Speaking Assessment",
    author = "Coulange, Sylvain  and
      Fries, Marie-H{\'e}l{\`e}ne  and
      Masperi, Monica  and
      Rossato, Solange",
    booktitle = "Proceedings of the 2024 Joint International Conference on Computational Linguistics, Language Resources and Evaluation (LREC-COLING 2024)",
    month = may,
    year = "2024",
    address = "Torino, Italia",

    url = "https://aclanthology.org/2024.lrec-main.27/",
    pages = "293--297"
}

@inproceedings{radford2023robust,
  title={Robust speech recognition via large-scale weak supervision},
  author={Radford, Alec and Kim, Jong Wook and Xu, Tao and Brockman, Greg and McLeavey, Christine and Sutskever, Ilya},
  booktitle={Proc. International Conference on Machine Learning (ICML)},
  pages={28492--28518},
  year={2023},
  organization={PMLR}
}

@book{mayer2003frog,
  title={Frog, where are you?},
  author={Mayer, Mercer},
  year={2003},
  publisher={Penguin}
}

\appendix

\section{Prompt for speechification}
\label{sec:appendix_speechify}

Below, we report the prompt used for GPT-4.1 \emph{speechification}, i.e., the transformation of essay-like texts into spoken-style learner transcripts. We highlight in bold the placeholders for the learner’s CEFR level and written response.

\begin{quote}

\texttt{Rewrite the following text as a spoken response by an English learner.}

\texttt{Speaker profile:}

\texttt{* Proficiency level (CEFR): \textbf{[CEFR LEVEL]}}

\texttt{Input text: \textbf{[WRITTEN RESPONSE]}}

\texttt{Core constraints:}

\texttt{* Keep the original meaning exactly the same.}

\texttt{* DO NOT correct grammar mistakes or improve the language.}

\texttt{* Preserve all learner errors.}

\texttt{* Do NOT introduce new factual content.}

\texttt{Goal:}

\texttt{Produce a realistic transcription of spontaneous speech, as if the learner is speaking and constructing the response in real time (not reading or summarising).}

\texttt{Discourse and structure:}

\texttt{* Convert the text into an incremental spoken narrative.}

\texttt{* Avoid compact or essay-like sentences.}

\texttt{* Use step-by-step development of ideas (as if recalling or explaining while speaking).}

\texttt{* Allow partial sentences, restarts, and local reformulations.}

\texttt{* Introduce mild redundancy (rephrasing, clarifying, or repeating ideas).}

\texttt{* Expand the text slightly to resemble spoken length (do not shorten).}

\texttt{Disfluency modelling (must be natural and controlled):}

\texttt{Include a mix of:}

\texttt{* Hesitations (e.g., ``um'', ``uh'')}

\texttt{* Repetitions (e.g., ``he-he goes'', ``they go, they go'')}

\texttt{* False starts and self-repairs (e.g., ``go-went'', ``to the- to the place'')}

\texttt{* Occasional word fragments (e.g., ``bi-bit'', ``f-falled'')}

\texttt{Placement rules:}

\texttt{* Insert disfluencies at natural planning points (before verbs, noun phrases, or difficult words).}

\texttt{* Include some mid-clause disfluencies, not only at sentence boundaries.}

\texttt{* Do NOT overuse any single type.}

\texttt{* Model disfluencies according to CEFR level.}

\texttt{* Model hesitation frequency according to CEFR level, distributing them naturally across the text (not clustered or periodic).}

\texttt{Cognitive/production effects:}

\texttt{Simulate real speaking behaviour:}

\texttt{* Include signs of uncertainty or memory limits (e.g., ``I think'', ``I don’t remember'', ``something like that'') when appropriate.}

\texttt{* Allow vague references (e.g., ``that thing'', ``that kind of things'').}

\texttt{* Do not make the speaker sound perfectly confident or precise.}

\texttt{* Avoid perfectly structured summaries.}

\texttt{Proficiency adaptation:}

\texttt{* Lower CEFR (A1–A2): shorter chunks, more hesitations, more repetitions, simpler structure.}

\texttt{* Mid CEFR (B1–B2): moderate disfluency, some repairs, longer sentences.}

\texttt{* High CEFR (C1–C2): mostly fluent, occasional hesitation for planning.}

\texttt{Formatting:}

\texttt{* Keep punctuation clear and readable.}

\texttt{* Use commas, dashes, and ellipses to reflect pauses and repairs.}

\texttt{* Do NOT produce raw ASR-style text.}

\texttt{* Output should remain compatible with standard written text processing.}

\texttt{Final requirement:}

\texttt{The output must resemble a real learner transcript, capturing both linguistic errors and the cognitive process of speaking, not just adding superficial noise.
Only output the transcript without adding any other comments or observations.}

\end{quote}

Here, we report the prompt used for speechification without CEFR-conditioning:

\begin{quote}

\texttt{Rewrite the following text as a spoken response by an English learner.}

\texttt{Input text: \textbf{[WRITTEN RESPONSE]}}

\texttt{Core constraints:}

\texttt{* Keep the original meaning exactly the same.}
\texttt{* DO NOT correct grammar mistakes or improve the language.}
\texttt{* Preserve all learner errors.}
\texttt{* Do NOT introduce new factual content.}

\texttt{Goal:}
\texttt{Produce a realistic transcription of spontaneous speech, as if the learner is speaking and constructing the response in real time (not reading or summarising).}

\texttt{Discourse and structure:}

\texttt{* Convert the text into an incremental spoken narrative.}
\texttt{* Avoid compact or essay-like sentences.}
\texttt{* Use step-by-step development of ideas (as if recalling or explaining while speaking).}
\texttt{* Allow partial sentences, restarts, and local reformulations.}
\texttt{* Introduce mild redundancy (rephrasing, clarifying, or repeating ideas).}
\texttt{* Expand the text slightly to resemble spoken length (do not shorten).}

\texttt{Disfluency modelling (must be natural and controlled):
Include a mix of:}

\texttt{* Hesitations (e.g., ``um'', ``uh'')}
\texttt{* Repetitions (e.g., ``he-he goes'', ``they go, they go'')}
\texttt{* False starts and self-repairs (e.g., ``go-went'', ``to the- to the place'')}
\texttt{* Occasional word fragments (e.g., ``bi-bit'', ``f-falled'')}

\texttt{Placement rules:}

\texttt{* Insert disfluencies at natural planning points (before verbs, noun phrases, or difficult words).}
\texttt{* Include some mid-clause disfluencies, not only at sentence boundaries.}
\texttt{* Do NOT overuse any single type.}

\texttt{Cognitive/production effects:}
\texttt{Simulate real speaking behaviour:}

\texttt{* Include signs of uncertainty or memory limits (e.g., ``I think'', ``I don’t remember'', ``something like that'') when appropriate.}
\texttt{* Allow vague references (e.g., ``that thing'', ``that kind of things'').}
\texttt{* Do not make the speaker sound perfectly confident or precise.}
\texttt{* Avoid perfectly structured summaries.}

\texttt{Formatting:}

\texttt{* Keep punctuation clear and readable.}
\texttt{* Use commas, dashes, and ellipses to reflect pauses and repairs.}
\texttt{* Do NOT produce raw ASR-style text.}
\texttt{* Output should remain compatible with standard written text processing.}

\texttt{Final requirement:}
\texttt{The output must resemble a real learner transcript, capturing both linguistic errors and the cognitive process of speaking, not just adding superficial noise.}
\texttt{Only output the transcript without adding any other comments or observations.}
\end{quote}

\end{document}